\newcommand{\ket}[1]{|{ #1}\rangle}
\newcommand{\bra}[1]{\langle{#1 }|}
\newcommand{\bk}{\bf k}
\newcommand{\br}{\bf r}
\newcommand{\fs}{$5\,\mathrm{S}_{1/2}$~}
\newcommand{\fpoh}{$5\,\mathrm{P}_{1/2}$~}
\newcommand{\fpth}{$5\,\mathrm{P}_{3/2}$~}
\newcommand{\rb}{$\,^{87}$Rb~}
\newcommand{\ds}{\displaystyle}

\documentclass[aps
        ,pra
        ,twocolumn
         ,superscriptaddress
         ,amsmath
]{revtex4}

\usepackage{color}
\usepackage{graphicx}
\usepackage{amssymb}

\begin{document}

\title{Theoretical analysis of the implementation of a quantum phase gate\\
       with neutral atoms on atom chips}

\author{E. Charron}
\affiliation{Laboratoire de Photophysique Mol\'eculaire du CNRS, B\^atiment 210 - Universit\'e Paris-Sud,
91405 Orsay Cedex - France}

\author{M. A. Cirone}
\affiliation{ECT*, Strada delle Tabarelle 286, I-38050 Villazzano, Trento, Italy, and \\
Dipartimento di Fisica, Universit\`a di Trento, and BEC-CNR-INFM, I-38050 Povo, Italy}

\author{A. Negretti}
\altaffiliation{Present address: Department of Physics and Astronomy, University of Aarhus, Ny Munkegade, Building 1520, DK-8000 Aarhus C, Denmark}
\affiliation{ECT*, Strada delle Tabarelle 286, I-38050 Villazzano, Trento, Italy, and \\
Dipartimento di Fisica, Universit\`a di Trento, and BEC-CNR-INFM, I-38050 Povo, Italy}
\affiliation{Institut f\"ur Physik, Universit\"at Potsdam, Am Neuen Palais 10, 14469 Potsdam, Germany}

\author{J.~Schmiedmayer}
\affiliation{Physikalisches Institut, Universit\"at Heidelberg,
69120 Heidelberg, Germany}

\author{T. Calarco}
\affiliation{ECT*, Strada delle Tabarelle 286, I-38050 Villazzano, Trento, Italy, and \\
Dipartimento di Fisica, Universit\`a di Trento, and BEC-CNR-INFM, I-38050 Povo, Italy}
\affiliation{ITAMP, Harvard Smithsonian Center for Astrophysics, and Department of Physics, Harvard University, Cambridge, MA 02138, USA}

\date{\today}

\begin{abstract}
We present a detailed, realistic analysis of the implementation of
a proposal for a quantum phase gate based on atomic vibrational
states, specializing it to neutral rubidium atoms on atom chips.
We show how to create a double--well potential with static
currents on the atom chips, using for all relevant parameters
values that are achieved with present technology. The potential
barrier between the two wells can be modified by varying the
currents in order to realize a quantum phase gate for qubit states
encoded in the atomic external degree of freedom. The gate
performance is analyzed through numerical simulations; the
operation time is $\sim 10$ ms with a performance fidelity above
$99.9$\%. For storage of the state between the operations the
qubit state can be transferred efficiently via Raman transitions
to two hyperfine states, where its decoherence is strongly
inhibited. In addition we discuss the limits imposed by the
proximity of the surface to the gate fidelity.
\end{abstract}

\maketitle

\section{Introduction}
\label{sec:Intro}

The idea of encoding information in quantum two--level systems
(qubits) instead of in classical bits promises a revolution in the
way we process and communicate information~\cite{niel}. Quantum
computers, i.e., processing units manipulating qubits instead of
classical bits, would lead to an intrinsic speed-up of calculation
that is not possible with a classical computer~\cite{shor,grov}.
For this purpose, a set of controlled operations is necessary,
that substitute the network of electronic logic gates of present
microelectronics. Analog to classical bits, also for qubits a
limited set of universal gates exists that allows to implement
networks for the execution of any quantum algorithm~\cite{gate}.
Such universal gates operate on single qubits and on pairs of
qubits. Whereas the design of single--qubit operations is
conceptually simple and experimental implementations are within
the reach of current technology in many cases, two--qubit
operations are more demanding in terms of both theoretical and
experimental investigations.

Several schemes for the realization of two--qubit gates with atoms
or ions have been proposed in the last
years~\cite{cira,turc,cala,jaks,bren,char} and a considerable
amount of progress has been recently realized for the future
implementation of quantum information with cold atoms in optical
lattices~\cite{mande,porto}. Many of these schemes have been
elaborated for ideal systems (harmonic oscillators, two--level
systems, etc.) or under simplifying approximations for the sake of
simplicity, in order to illustrate the idea on which the gate is
based. The realization of such schemes in an experimental setup,
where the characteristics of the real physical system are
inevitably to be taken into account, is often a formidable task.
The inclusion of realistic features and values requires a
reexamination and modification of the schemes presented in the
literature, in order to exploit all properties of the real
physical system. For instance, the collisional gate with switching
potentials proposed in \cite{cala} exploits the complete revivals
of wave packets in harmonic traps, which are difficult to achieve
for neutral atoms, and deviations from harmonicity pose limiting
restrictions to the performance of the gate \cite{negr}. More
realistic proposals for quantum gates are thus required.

With the spirit just outlined, we reconsider the two--qubit gate
initially proposed in Ref.~\cite{char}. Two atoms are trapped in a
double well potential. The ground ($\ket{g}$) and first excited
($\ket{e}$) vibrational states of each well are used as qubit
logic states and the phase gate operation occurs via selective
interaction between the atoms in the classically forbidden region
under the potential barrier that separates the two atoms. In this
paper we investigate the feasibility of such a scheme with neutral
atoms on atom chips. Instead of assuming ideal trapping potentials
for the atoms to facilitate numerical analysis, we derive here the
exact potential created by appropriate stationary and
time-dependent currents and bias magnetic fields on atom chips.
The finite size of the current--carrying wires is also taken into
account and realistic values for all physical parameters are used.
In addition we consider also effects of the proximity of the
surface, like spin flips and decoherence. Moreover, we specialize
our discussion to neutral \rb atoms. In this way our
investigations are much closer to the experimental conditions and
our results show how the implementation of such a scheme is indeed
feasible.

In spite of the improved realism of our investigations, some
features of the real system are neglected in our analysis.
Nonetheless, we expect that their role is marginal and/or can be
minimized. Van deer Waals and Casimir forces between the atoms and
the chip surface are not taken into account, since these effects
become negligible when the atoms are kept sufficiently distant
from the chip surface. Similarly, fabrication defects, such as
roughness of the chip surface and imperfections in the atom chip
are not taken into account, since they were shown to be
sufficiently small, and the potentials sufficiently smooth
\cite{kruger05} when using the appropriate fabrication techniques
\cite{Gro04}. Thermal fluctuations of currents in the metal layers
of the atom chip are another important issue \cite{Henkel,Rekdal04,Varpula84}
They may lead to loss of the
atoms, and to decoherence of quantum states. However, we shall
show that one can transfer the qubit state from the external
degree of freedom to the internal one, using two clock states
whose decoherence is strongly reduced~\cite{harb,treu} and the important aspect
to consider is the atom loss. Therefore, in
spite of these approximations, our analysis is a significant step
towards the first implementation of a phase gate with neutral
atoms on atom chips.


The outline of the paper is as follows: in Sec.~\ref{sec:Pot} we
discuss the properties of \rb atoms that are relevant for our
analysis and describe the architecture of wires and bias magnetic
fields on atom chips that create a double--well potential for the
two atoms with strong confinement in one dimension. We also give
the values of currents, bias fields, and wire sizes that realize
this trap. In Sec.~\ref{sec:Gate} we give the results of our
numerical investigations on the performance of the phase gate. We
show that high fidelity ($\geqslant 99.9\%$) can be achieved with
short operation times ($\sim 10$~ms).  In Sec.~\ref{sec:rama} we
show how to transfer the qubit state from the external degree of
freedom to two hyperfine (magnetic tappable clock) states with
Raman transitions. The relative phase of the two hyperfine states
is insensitive, to first order, to fluctuations of the magnetic
fields, thus, preserving the purity of the qubit state for long
time. Finally in Sec.~\ref{sec:loss} we relate the achievable gate
operation times to the expected atomic life times and coherence
time, to estimate a realistic fidelity.  Whereas the decoherence
can be avoided by storing the qubits in the clock states, the loss
of qubits due to thermally induced spin flips remains, and limits
the fidelity of operations. The Appendix contains details
concerning the magnetic field created by the currents on the atom
chip.

\section{Double-well potentials with magnetic fields on atom chips}
\label{sec:Pot}

Atom chips are versatile integrated microstructures for the
manipulation of samples of atoms in the ultracold and quantum
degenerate regime (see Ref.~\cite{folm} and references therein). Trapping potentials for
neutral atoms can be created near the chip surface. We shall
consider the simple case of homogeneous bias magnetic fields and
magnetic fields created by dc (but time--dependent) currents. The spin of slow, cold atoms
remains adiabatically aligned with the magnetic field. The trapping
magnetic potential, in the weak field approximation, is expressed
by
\begin{equation}
V({\bf r})=-g_F \mu_B m_F B({\bf r})
\label{magp}
\end{equation}
where $\mu_B$ is the Bohr magneton, $g_F$ is the Land\'e factor,
$m_F$ is the azimuthal quantum number, and $B({\bf r})$
is the magnetic field. In section~\ref{sec:Gate},
a higher order accurate formula --~derived from the Breit-Rabi
approach~\cite{breit}~-- will be used in the numerical simulations,
but we limit our discussion to the first order term given in
Eq.~(\ref{magp}) for the sake of simplicity.

\begin{figure}[t!]
\centering
\includegraphics[width=8.6cm,angle=0,clip]{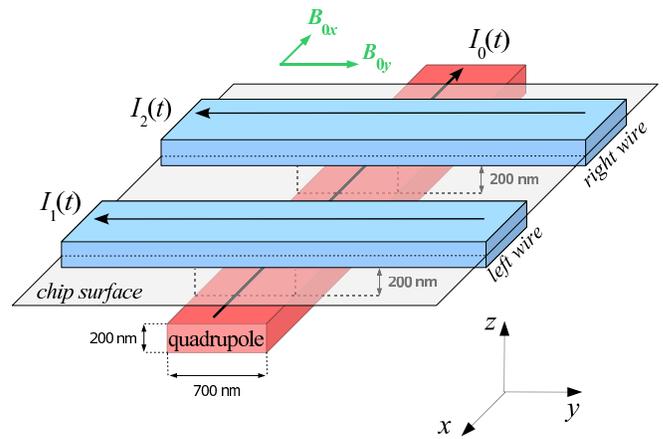}
\caption{(Color online) Schematic view of the atom chip configuration.
The two wires along the $y$--axis lie on the chip surface and are
separated from the quadrupole wire by 200~nm.}
\label{fig:wires}
\end{figure}

The most simple configuration of wires and bias magnetic fields
that creates a double--well potential is shown in
Fig.~\ref{fig:wires} \cite{homm}. A longitudinal wire along $x$
(hereafter, quadrupole wire) carrying a dc current $I_0$ and a
uniform bias magnetic field $B_{0y}$ perpendicular to the wire
create a quadrupole potential, with a zero magnetic field along a
line parallel to the quadrupole wire. Clearly, along this line the
magnetic field is minimum, but Maiorana spin flips occur at zero
magnetic field, which result in atom losses. Therefore the minimum is
shifted to a non--zero value with the
addition of a second uniform bias magnetic field $B_{0x}$,
orthogonal to the first one and parallel to the chip surface. Two
more wires (hereafter, left and right wire, respectively),
perpendicular to the quadrupole wire, carry a dc current
\mbox{$I_{1,2} = \alpha I_0$}, whose magnetic fields give rise to
a modulation of the trapping potential.

In consideration of the small extension of the system (a few
microns), the assumption of infinitely thin wires is too crude
an approximation. Therefore in the potential~(\ref{magp}) we shall
use the expression of the magnetic field created by currents
flowing through wires with finite rectangular sections (a typical
shape of wires on atom chips). Details of the calculation are
given in the Appendix.

\begin{figure}[t!]
\centering
\includegraphics[width=8.6cm,angle=0,clip]{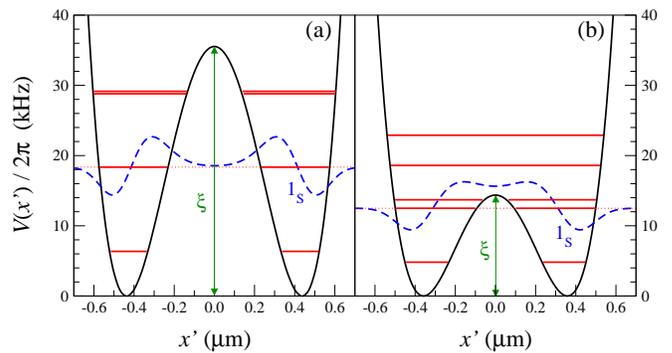}
\caption{(Color online) Double well potentials created by the atom chip
configuration shown in Figure~\ref{fig:wires}. The energies of the first
six eigenstates are shown as red horizontal lines. The blue dashed line
represents the wavefunction of the third eigenstate labeled as
$1_{\mathrm{S}}$ because it originates from the symmetric combination
of the $v=1$ trapped levels, also labeled as $\ket{e}$ in the text.
(a)~Highest barrier $\xi/2\pi=35.4$~kHz obtained with $I_0=40.89$~mA
and $\alpha=70.25\times10^{-3}$. (b)~Lowest barrier $\xi/2\pi=14.4$~kHz
obtained with $I_0=42.01$~mA and $\alpha=69.70\times10^{-3}$. In both
cases the bias magnetic fields are equal to $B_{0x}=-9.90$~G, and
$B_{0y}=50.0$~G.}
\label{fig:potentials}
\end{figure}

A trapping potential with two well-separated minima, as shown in
Figure~\ref{fig:potentials}(a), is created with the values
\mbox{$I_0=40.89$~mA}, \mbox{$\alpha=70.25 \times 10^{-3}$},
\mbox{$B_{0x}=-9.90$~G}, and \mbox{$B_{0y}=50.0$~G}. The three wires have an
identical rectangular section, of width \mbox{$W=700$~nm} and height
\mbox{$H=200$~nm}, and the centers of the left and right wires are
\mbox{$1.60~\mu$m} apart. The center of the quadrupole wire is at a
distance \mbox{$z_Q=400$~nm} under the chip surface, whereas the left
and right wires lie on the chip. We stress that the values
for the required transversal confinement has been approached even in the first experiments
with nano fabricated atom chips \cite{fol00,folm} and the currents, bias fields, size and distances of the wires
are within current laboratory conditions \cite{notedens}.

The two potential minima are at a distance of 1.19~$\mu$m from
the surface, and the line joining them is slightly tilted by an
angle \mbox{$\beta \simeq 14.8^\circ$} from the $x$ axis. This angle
defines the new axis $x'$ along which the dynamics will take place
(see~\cite{ciro} for details). We also define a new axis $y'$
parallel to the chip surface and perpendicular to $x'$. The
$z$ axis remains unchanged. Since the trapping frequencies at the
two minima verify the condition \mbox{$\nu_{x'} \ll \nu_{y'} \simeq
\nu_{z}$}~\cite{notefreq}, transverse vibrations do not get excited during the
gate operation (provided that the energy of the atoms is less than the quanta of
transverse vibrations), and the trapped atoms will experience a quasi one--dimensional (1D) dynamics.

The value of the magnetic field at the two minima is \mbox{$B_m \simeq 3.23$~G}.
This value minimizes the decoherence induced by fluctuations of the dc
currents for the hyperfine states $\ket{F=2,m_F=1}$ and
$\ket{F=1,m_F=-1}$ of the ground state \fs of \rb~\cite{treu}.
These clock states will be used to store the qubit information at the end of the
gate operation, as described in Sec.~\ref{sec:rama}.

\section{Gate operation}
\label{sec:Gate}

Compared to previous static proposals~\cite{ciro}, we implement here a
time-dependent phase gate in a spirit similar to the one of
Ref.~\cite{char}.

As it can be noticed in Figure~\ref{fig:potentials}(a), when the barrier
is high the translational wavefunctions of the atoms do not overlap in
the inter--well region. In this type of environment, the atoms do not
interact. On the other hand, when the barrier is lowered, as in
Figure~\ref{fig:potentials}(b), tunneling takes place and the probability
of finding the atoms in the classically forbidden region is not negligible
any more. As a consequence, the energy splitting between the symmetric and
anti--symmetric state combinations increases quickly when the barrier height
$\xi$ is lowered. This effect is clearly state-selective in the sense that it
affects differently the ground and excited translational states. It
therefore constitutes an interesting candidate for the implementation
of a conditional logical gate.

\begin{figure}[t!]
\centering
\includegraphics[width=8.6cm,angle=0,clip]{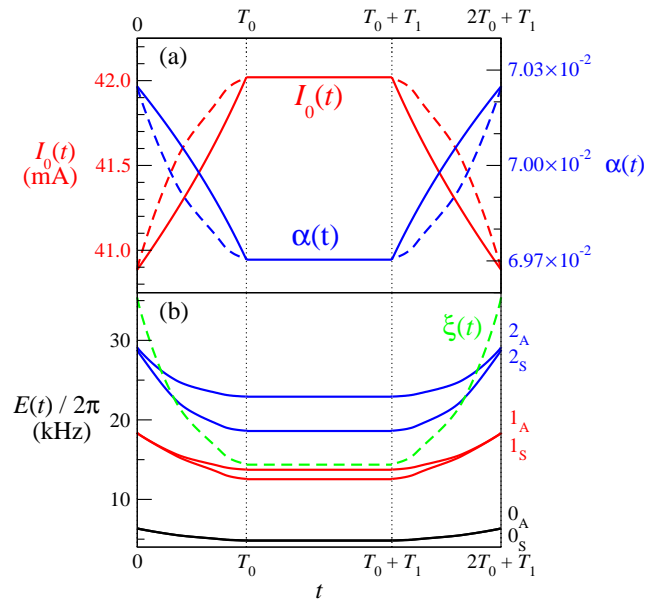}
\caption{(Color online) (a) Variation of the dc current $I_0(t)$ (red lines)
in the quadrupole wire and of \mbox{$\alpha(t)=I_1(t)/I_0(t)=I_2(t)/I_0(t)$}
(blue lines) during the gate operation. The solid lines correspond to a
simple linear gate and the dashed lines represent an optimized gate (see text
for details). (b) Variation of the barrier height $\xi(t)$ (green dashed
line) and of the energies of the first six instantaneous eigenstates of the
double well potential (solid lines) in the case of the optimized gate. The times
$T_0$, $(T_0+T_1)$ and $(2T_0+T_1)$ delimit the three steps which constitute
the conditional $\pi$ phase gate.}
\label{fig:gate}
\end{figure}

In the present scheme, the barrier height $\xi(t)$ is controlled by
varying simultaneously the intensities $I_0(t)$ and \mbox{$I_1(t) = I_2(t)
= \alpha(t) \times I_0(t)$} in the quadrupole and in the perpendicular wires.
In a first and simple implementation of the phase gate, we impose a linear variation of
the barrier height $\xi$ with time. The phase gate is decomposed
in three steps:
\begin{itemize}
\item
When \mbox{$0 \leqslant t \leqslant T_0$} the barrier is lowered and the
double--well potential changes from the one of Figure~\ref{fig:potentials}(a)
to the one of Figure~\ref{fig:potentials}(b).
\item
When \mbox{$T_0 \leqslant t \leqslant T_0+T_1$} the inter--well barrier is
fixed at its lowest value, such that a large inter--atomic interaction takes
place.
\item
Finally, when \mbox{$T_0+T_1 \leqslant t \leqslant 2T_0+T_1$} the inter--well
barrier is raised again until the initial condition is recovered.
\end{itemize}
The linear variation of $\xi(t)$ is obtained by changing $I_0(t)$ and
$\alpha(t)$ only, as depicted by the solid lines shown in Figure~\ref{fig:gate}(a).
We have verified that this simultaneous variation of the dc currents does not
modify the direction $x'$ along which the dynamics is taking place. In addition,
the value of the magnetic field at the potential minima remains equal to 3.23~G
during the whole gate operation. The dynamics is therefore quasi--adiabatic if
\mbox{$T_0 \gg 1/\nu_{x'} \simeq 77~\mu$s} (hereafter we write $x,y$ instead of $x',y'$).

The gate operation is followed by solving the time--dependent Schr\"odinger
equation along the double-well direction $x$ for the wave packet
$\Psi(x_1,x_2,t)$ describing the motion of the two atoms
\begin{equation}
\label{TDSE}
i\hbar\,\frac{\partial}{\partial t}\,\Psi(x_1,x_2,t) = {\cal H}(x_1,x_2,t)\,\Psi(x_1,x_2,t)\,.
\end{equation}

\begin{figure}[t!]
\centering
\includegraphics[width=8.6cm,angle=0,clip]{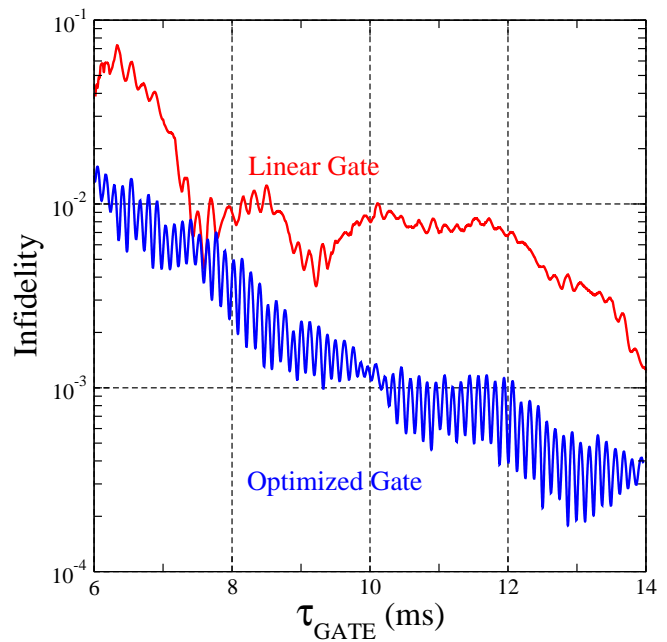}
\caption{(Color online) Infidelity of the conditional phase gate as a
function of the gate duration. The red and blue lines correspond to the
infidelities calculated for the linear and optimized gates respectively
(see Figure~\ref{fig:gate} and text for details).}
\label{fig:infidelity}
\end{figure}

The two--dimensional time--dependent Hamiltonian is written as
\begin{equation}
\label{H2D}
{\cal H}(x_1,x_2,t)=\hat{T}_{x_1}+\hat{T}_{x_2}+V_{\mathrm{2D}}(x_1,x_2,t)\,,
\end{equation}
where $\hat{T}_{q}$ denotes the kinetic energy operator along the
$q$--coordinate. The two--dimensional potential is given by the
following sum
\begin{eqnarray}
V_{\mathrm{2D}}(x_1,x_2,t) & = & V(x_1,t) + V(x_2,t)\nonumber\\
& & +\;V_{\mathrm{int}}(|x_2-x_1|,t)\,,
\label{V2d}
\end{eqnarray}
where $V(x,t)$ is the trapping potential~(\ref{magp}) created by
the atom chip and $V_{\mathrm{int}}(|x_2-x_1|,t)$ represents the averaged
interaction potential between the two atoms at time $t$
\begin{equation}
V_{\mathrm{int}}(|x_2-x_1|,t) = 2 \hbar a_0
\left(\omega_{y}\,\omega_{z}\right)^{1/2}
\delta\left(|x_2-x_1|\right)\,.
\label{Vint}
\end{equation}
This last expression is obtained by averaging the three--dimensional
delta function interaction potential over the lowest trap states along
the $y$ and $z$ directions~\cite{cala}. One can note that the atom--atom
interaction strength is proportional to the s--wave scattering length
$a_0$. Since the orthogonal trapping frequencies $\omega_{y}$ and
$\omega_{z}$ vary slightly during the gate operation~\cite{notefreq},
the averaged interaction strength is also slightly time-dependent.

We solve the time-dependent Schr\"odinger equation~(\ref{TDSE}) in a basis set
approach by propagating the initial state of the two--atom system in time. We
start with the atoms initially in one of the first four eigenstates $\ket{gg}$,
$\ket{ge}$, $\ket{eg}$, or $\ket{ee}$ of the double--well potential~(\ref{V2d}).
The wavefunction $\Psi(x_1,x_2,t)$ is then expanded as
\begin{equation}
\Psi(x_1,x_2,t) = \sum_i c_i(t) \times \varphi_i(x_1,x_2,t)\,
\label{PsiExp}
\end{equation}
where $\varphi_i(x_1,x_2,t)$ represent the wavefunctions associated to the
instantaneous eigenstates of the two--dimensional potential $V_{\mathrm{2D}}(x_1,x_2,t)$
of Eq.~(\ref{V2d}). Inserting this expansion into the time--dependent Schr\"odinger
equation~(\ref{TDSE}) yields the following set of first order coupled ordinary
differential equations for the complex coefficients $c_i(t)$
\begin{equation}
i\hbar\,\frac{d}{dt}c_i(t) = \varepsilon_{i}\,c_i(t) -i\hbar \sum_j c_j(t)\,V_{ij}(t)\,,
\label{ode}
\end{equation}
where $\varepsilon_{i}$ denotes the energy of the eigenstate $\varphi_i$ and $V_{ij}(t)$
is a time--dependent non--adiabatic coupling arising from the time variation of the barrier
height $\xi(t)$
\begin{equation}
V_{ij}(t) = \left<\varphi_i\right|\frac{\partial}{\partial t}\left|\varphi_j\right>
= \frac{d\xi}{dt}\,
\left<\varphi_i\right|\frac{\partial}{\partial\xi}\left|\varphi_j\right>\,.
\label{nac}
\end{equation}
This set of equations is solved using an accurate Shampine--Gordon
algorithm~\cite{shamp}. For each value of the duration $T_0$ we adapt $T_1$ in order to
obtain an accurate conditional phase gate, with $\phi=\pi \pm 10^{-6}$.

At the end of the propagation ($t=t_f$) the coefficients $c_i(t_f)$ are
analyzed to calculate the infidelity of the gate
\begin{equation}
I = \sum_{i = \ket{gg}\cdots\ket{ee}} \left(1-\left|c_{i}(t_f)\right|^2\right)\,.
\label{infidelity}
\end{equation}
where the sum runs over all possible initial qubit states.
The infidelity is therefore a measure of the deviation from adiabaticity
which arises from the non-adiabatic couplings $V_{ij}(t)$. This quantity
is plotted in Figure~\ref{fig:infidelity} (red line for the linear gate)
as a function of the gate duration. It shows an oscillatory behavior
partially similar to the one observed with atoms trapped in an optical
lattice~\cite{char}. The succession of maxima and minima is a signature
of constructive and destructive interferences between two distinct
pathways of excitation of the initial qubit state. Indeed the initial
state may be excited in the time intervals \mbox{$0 \leqslant t \leqslant
T_0$} and \mbox{$T_0+T_1 \leqslant t \leqslant 2T_0+T_1$}, when
the barrier is lowered and raised. The nature of this interference depends
on the phases which develop during the gate operation~\cite{char}.
The periodicity of the oscillation is simply related to the Bohr frequencies
associated with the energy splitting of the two--atom eigenstates. The
linear gate configuration proposed here can achieve a relatively high
fidelity of about 99.6\% in just 7.6~ms.

One should also realize that in the general case the couplings between the
initial qubit states and the other accessible two--atom eigenstates vary
with time. These couplings effectively increase when the inter--well barrier
approaches the energy of the initial state. The linear gate proposed until
now is therefore far from being optimal for the maximization of the gate
fidelity.

We have thus implemented an optimized gate which tends to minimize
these couplings during the whole gate duration. For this purpose, we impose
a fast variation of the barrier height $\xi(t)$ at early times \mbox{$t
\ll T_0$}, whereas this variation is much slower when \mbox{$t \simeq T_0$}.
This is done by choosing
\begin{equation}
\hbar\frac{d\xi}{dt} = \gamma \mathrm{Min}_{\,i\,} \left|\frac{\varepsilon_i-\varepsilon_{\ket{ee}}}
{\ds\left<\varphi_i\right|\frac{\partial}{\partial\xi}\left|\varphi_{\ket{ee}}\right>}\right|\,.
\label{opt}
\end{equation}
In this expression, $\gamma$ is a dimensionless proportionality factor, which can
be decreased to achieve larger gate durations. The first derivative with respect
to time of the barrier height $\xi(t)$ is therefore chosen such that the maximum
effective first--order coupling between the highest energy qubit state $\ket{ee}$
and all other states remains constant during the whole gate duration. With this
approach, higher fidelities are expected when compared to a linear gate of the
same duration.

The variations of $I_0(t)$ and $\alpha(t)$ for this optimized gate are shown as
dashed lines in Figure~\ref{fig:gate}(a). Figure~\ref{fig:infidelity} shows that
the optimized gate infidelity (blue line) is, on average, improved by a factor
6 when compared to the linear gate. As a consequence, this optimized gate can
achieve fidelities of 99\% in 6.3~ms and of 99.9\% in 10.3~ms.

\section{Using both degrees of freedom as qubit states}
\label{sec:rama}

When neutral atoms are used, it is possible to employ either the
vibrational or the internal states as qubit states.
Section~\ref{sec:Gate} has shown that vibrational states are very
promising in terms of gate performance. However, the readout of
the qubit after the gate operation seems to be a difficult
task in comparison to the readout of internal states,
for which observation of fluorescence emitted from selected
transitions is a well established technique.
Moreover, a closer examination of the hyperfine structure of
the ground state of \rb provides further motivation for
the use of internal states.

\begin{figure}[t!]
\centering
\includegraphics[width=8.6cm,angle=0,clip]{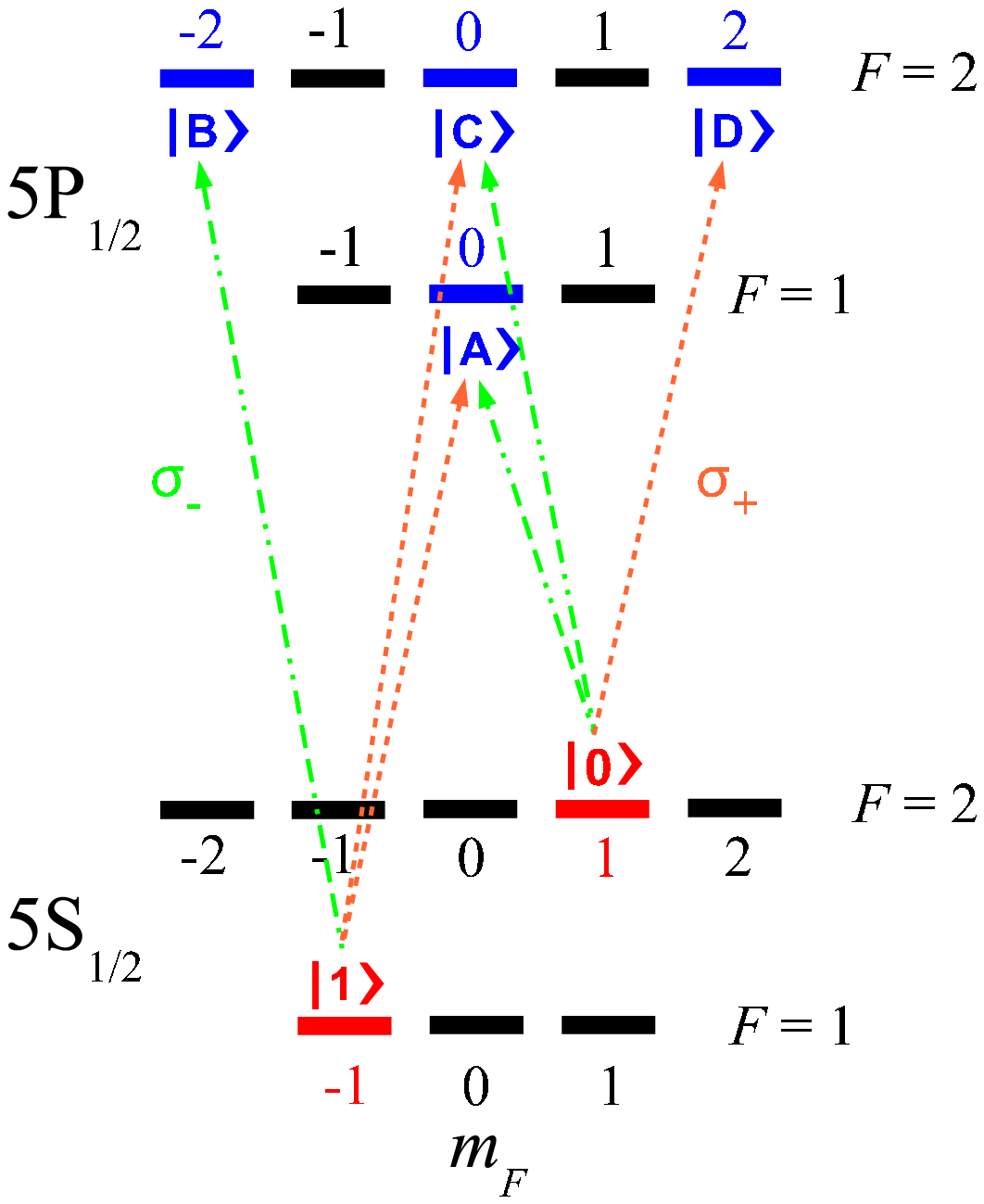}
\caption{(Color online) Schematic representation of the hyperfine
structure of the \fs and \fpoh states of \rb. The two qubits
$\ket{0}\equiv \ket{F=2,m_F=1}$ and $\ket{1}\equiv \ket{F=1,m_F=-1}$
of the \fs ground state are shown in red. The intermediate sublevels
$\ket{\rm A}$, $\ket{\rm B}$, $\ket{\rm C}$, and $\ket{\rm D}$ of the \fpoh excited
state are shown in blue. The $\sigma_{+}$ and $\sigma_{-}$ transitions
connecting them are shown with dotted (orange) and dash-dotted (green) arrows respectively.}
\label{fig:levels}
\end{figure}

The hyperfine structure of the \fs ground state of \rb is shown in Figure~\ref{fig:levels}:
eight sublevels have to be considered, three with $F=1$ and five with $F=2$.
Only three of these eight sublevels (the states $\ket{F=1,m_F=-1}$, $\ket{F=2,m_F=1}$, and
$\ket{F=2,m_F=2}$) are low--field seeking states which can be trapped by
static magnetic potentials~\cite{folm}. In addition, the two hyperfine
levels $\ket{F=1,m_F=-1}$ and $\ket{F=2,m_F=1}$ have opposite Land\'e factors. As a
consequence, they experience identical magnetic potentials. A more precise estimation
that goes beyond the linear approximation given in Eq.~(\ref{magp}) shows
that the difference of the Zeeman shifts experienced by these two states is
minimum at \mbox{$B \simeq 3.23$~G~\cite{harb}}. At this field the states $\ket{0}\equiv
\ket{F=2,m_F=1}$ and $\ket{1}\equiv \ket{F=1,m_F=-1}$ are insensitive to field fluctuations
at first order. The decoherence of an arbitrary superposition of the two states $\ket{0}$
and $\ket{1}$ due to fluctuations in the current intensities (which entail fluctuations
in the magnetic field and thus in the potential~(\ref{magp})) is thus strongly
inhibited. Indeed, coherent oscillations between the states $\ket{0}$ and $\ket{1}$ have
recently been observed and the decoherence time has been estimated to be
\mbox{$\tau_c=2.8$~s~$\pm 1.6$~s~\cite{treu}}. The inhibition of decoherence is a fundamental
issue in quantum computation, since the coherence of the quantum state is an
essential ingredient. It would thus be desirable to use the clock states $\ket{F=1,m_F=-1}$
and $\ket{F=2,m_F=1}$ for the storage of quantum information.

In fact, we can use both degrees of freedom as qubits: when
information must be processed, it will be encoded in the external
degree of freedom and the gate operation will be done as described in
the previous Section. At the end of the gate operation, the qubit
state can be transferred to the internal degree of freedom and
stored there until new operations are performed or until the final
readout. In order to realize this transfer from one degree of
freedom to the other, we need to swap the states associated with these
two degrees of freedom, according to
\begin{eqnarray}
\ket{g1} \leftrightarrow \ket{e0}
\label{map}
\end{eqnarray}
In this way, an arbitrary qubit state encoded in the vibrational states as
\begin{equation}
\label{qubitintial}
\left( a\ket{gg}+b\ket{ge}+c\ket{eg}+d\ket{ee} \right) \otimes \ket{00}
\end{equation}
is transformed into
\begin{equation}
 \ket{gg} \otimes \left( a\ket{00}+b\ket{01}+c\ket{10}+d\ket{11} \right)
\end{equation}
and vice versa. Eq.~(\ref{qubitintial}) assumes that the
internal degree of freedom of the two atoms has been initialized to
the state $\ket{0}$. The map given in Eq.~(\ref{map})
is a selective switch from the internal state $\ket{0}$ (
respectively $\ket{1}$) to the state $\ket{1}$ (
respectively $\ket{0}$) under the control of the vibrational state.
In order to realize it, we propose the use of two--photon Raman
transitions.

The realization of such a two--photon Raman transition via
intermediate $5\,\mathrm{P}$ states requires a careful analysis~\cite{stec}.
The \fs and \fpoh levels have eight hyperfine states with $F=1$ or 2,
whereas the \fpth level comprise sixteen states with $F=0$, 1,
2 or 3. The Raman transition requires two lasers, one with $\sigma^+$
polarization (frequency $\omega_+$ and wave vector $\mathbf{k}_+$), the
other with $\sigma^-$ polarization (frequency $\omega_-$ and wave vector
$\mathbf{k}_-$). When the laser frequencies $\omega_+$ and
$\omega_-$ are kept below the \fs$\rightarrow$\fpoh transition, the states
in \fpth can be neglected, being detuned by about 7~THz. From now on we therefore
restrict our analysis to the states $\ket{0}$ and $\ket{1}$, and to the following \fpoh states
\begin{displaymath}
\begin{array}{rcccrcc}
\ket{\rm A} & \equiv & \ket{F=1,m_F=0}  & ~ & \ket{\rm C} & \equiv & \ket{F=2,m_F=0} \\
\ket{\rm B} & \equiv & \ket{F=2,m_F=-2} & ~ & \ket{\rm D} & \equiv & \ket{F=2,m_F=2}
\end{array}
\end{displaymath}
Taking into account all possible transitions according to the selection
rules imposed by the polarizations, the Hamiltonian in the
usual rotating--wave approximation is
\begin{equation}
\mathcal{H} =\mathcal{H}_0+\mathcal{H}_1+\mathcal{H}_2
\label{hami}
\end{equation}
with
\begin{eqnarray*}
\mathcal{H}_0 & = & \hbar\sum_{k=0,1}\omega_{k}\,\ket{k}\bra{k}+\hbar
\sum_{j={\rm A}}^{\rm D}\omega_{j}\ket{j}\bra{j} \\
\mathcal{H}_1 & = & \frac{\hbar\,\Omega_{+}}{2\sqrt{3}}\;e^{i\,\eta_+}
\Big(\ket{\rm A}\bra{1}+\ket{\rm C}\bra{1}-\sqrt{2}\,\ket{\rm D}\bra{0}\Big)+{\rm h.c.}\\
\mathcal{H}_2 & = & \frac{\hbar\,\Omega_{-}}{2}\;e^{i\,\eta_-}
\Big(\sqrt{2}\,\ket{\rm B}\bra{1}+\ket{\rm C}\bra{0}-\ket{\rm A}\bra{0}\Big)+{\rm h.c.}
\end{eqnarray*}
where we have defined the phases
\begin{displaymath}
\left\{
\begin{array}{ccc}
\eta_{+} & = & {\bk_+\cdot\br}-\omega_+\,t\\
\eta_{-} & = & {\bk_-\cdot\br}-\omega_-\,t
\end{array}
\right.
\end{displaymath}
and where $\Omega_{+}$ and $\Omega_{-}$ are the Rabi frequencies associated
with the interaction between the electric dipole of the atom and the electric field
of the lasers. $\br$ is the atomic position operator, and h.c. denotes the Hermitian
conjugate.

From the Hamiltonian~(\ref{hami}) one can derive the Heisenberg equations
of motion for the operators \mbox{$\sigma_{ij}\equiv\ket{i}\bra{j}$}, with \mbox{$i,j=0$,}
1, A, $\ldots$, D. For instance, for the operator $\sigma_{{\rm A}1}$ we get
\begin{eqnarray}
\label{eq:heisigma}
i\,\hbar\,\dot{\sigma}_{{\rm A}1} & = & \left[\sigma_{{\rm A}1},\mathcal{H}\right]\nonumber\\
& = & \left[\sigma_{{\rm A}1},\mathcal{H}_0\right]+\left[\sigma_{{\rm A}1},\mathcal{H}_1\right]
+ \left[\sigma_{{\rm A}1},\mathcal{H}_2\right]
\end{eqnarray}
where
\begin{eqnarray*}
\left[\sigma_{{\rm A}1},\mathcal{H}_0\right] & = &
\hbar\,\left(\omega_1-\omega_{{\rm A}}\right)\,\sigma_{{\rm A}1}\\
\left[\sigma_{{\rm A}1},\mathcal{H}_1\right] & = &
\frac{\hbar\,\Omega_{+}}{2\sqrt{3}}\;e^{-i\,\eta_+}\,
\Big(\sigma_{{\rm A}{\rm A}}-\sigma_{11}+\sigma_{{\rm A}{\rm C}}\Big)\\
\left[\sigma_{{\rm A}1},\mathcal{H}_2\right] & = &
\frac{\hbar\,\Omega_{-}}{2}\;e^{-i\,\eta_-}\,
\Big(\sigma_{01}+\sqrt{2}\,\sigma_{{\rm A}{\rm B}}\Big)\,.
\end{eqnarray*}

With the introduction of the ``slow'' variables
\begin{eqnarray*}
\tilde{\sigma}_{{\rm A}1} & \equiv & \sigma_{{\rm A}1}\;\;e^{-i\,\omega_+\,t}\\
\tilde{\sigma}_{01}       & \equiv & \sigma_{01}\;\;e^{-i\,(\omega_+-\omega_-)\,t}\,,
\end{eqnarray*}
equation~(\ref{eq:heisigma}) becomes
\begin{eqnarray*}
i\,\dot{\tilde{\sigma}}_{{\rm A}1} & = & \left(\omega_+-\omega_{{\rm A}}+\omega_1\right)\,
\tilde{\sigma}_{{\rm A}1} \\
& + & \frac{\Omega_{+}}{2\sqrt{3}}\;e^{-i\,\bk_+\cdot\br}\;
\Big(\sigma_{{\rm A}{\rm A}}-\sigma_{11} +\sigma_{{\rm A}{\rm C}}\Big) \\
& + & \frac{\Omega_{-}}{2}\;e^{-i\,\bk_-\cdot\br}\;
\Big(\tilde{\sigma}_{01}+\sqrt{2}\,e^{i\,(\omega_-\omega_+)\,t}\,\sigma_{{\rm A}{\rm B}}\Big)
\end{eqnarray*}
When the lasers are strongly detuned from the atomic transition frequencies, one can
obtain an effective Hamiltonian by imposing the adiabaticity condition
\mbox{$\dot{\tilde{\sigma}}_{{\rm A}\,1} \simeq 0$}~\cite{eber,wang}. In this case, we get
\begin{eqnarray*}
\sigma_{{\rm A}1} & = & \frac{\Omega_{+}}{2\sqrt{3}}\;\frac{e^{-i\,\eta_+}}{\Delta_{{\rm A}1}}\;
                        \Big(\sigma_{{\rm A}{\rm A}}-\sigma_{11}+\sigma_{{\rm A}{\rm C}}\Big) \\
                  &   & +\;\frac{\Omega_{-}}{2}\;\frac{e^{-i\,\eta_-}}{\Delta_{{\rm A}1}}\;
                        \Big(\sigma_{01}+\sqrt{2}\,\sigma_{{\rm A}{\rm B}}\Big)\,,
\end{eqnarray*}
where we have defined the detuning
\begin{equation}
\label{det}
\Delta_{{\rm A}1} \equiv \omega_{{\rm A}}-\omega_1-\omega_+\,.
\end{equation}
Performing similar calculations for the operators $\sigma_{{\rm B}1}$, $\sigma_{{\rm C}1}$,
$\sigma_{{\rm A}0}$, $\sigma_{{\rm C}0}$ and $\sigma_{{\rm D}0}$ after substituting them
into $\mathcal{H}_1$ and $\mathcal{H}_2$ and including the diagonal terms into $\mathcal{H}_0$,
we get the effective interaction Hamiltonians
\begin{equation}
\begin{array}{cccl}
\tilde{\mathcal{H}}_1 & = & \ds\frac{\hbar\,\Omega_{+}}{2\sqrt{6}}\,\Big[ &
                            \ds\Omega_{-}\;e^{i\,\eta}
                            \left(\frac{\sigma_{{\rm A}{\rm B}}}{\Delta_{{\rm A}1}}\right)+
                            \frac{\Omega_{+}}{\sqrt{6}}
                            \left(\frac{\sigma_{{\rm A}{\rm C}}}{\Delta_{{\rm A}1}}+
                                  \frac{\sigma_{{\rm A}{\rm C}}}{\Delta_{{\rm C}1}}\right)\\
& & & \ds +\;\Omega_{-}\;e^{-i\,\eta}
      \left(\frac{\sigma_{{\rm A}{\rm D}}}{\Delta_{{\rm D}0}}+
      \frac{\sigma_{{\rm B}{\rm C}}}{\Delta_{{\rm C}1}}-
      \frac{\sigma_{{\rm C}{\rm D}}}{\Delta_{{\rm D}0}}\right)\\
& & & \ds +\;\frac{\Omega_{-}}{\sqrt{2}}\;e^{i\,\eta}
      \left.\left(\frac{\sigma_{01}}{\Delta_{{\rm A}1}}-
                  \frac{\sigma_{01}}{\Delta_{{\rm C}1}}\right)\,\right]+\rm h.c.
\end{array}
\end{equation}
and
\begin{equation}
\begin{array}{cccl}
\tilde{\mathcal{H}}_2 & = & \ds\frac{\hbar\,\Omega_{-}}{2\sqrt{2}}\,\Big[ &
                            \ds\frac{\Omega_{+}}{\sqrt{3}}\;e^{i\,\eta}
                            \left(\frac{\sigma_{{\rm A}{\rm B}}}{\Delta_{{\rm B}1}}\right)-
                            \frac{\Omega_{-}}{\sqrt{2}}
                            \left(\frac{\sigma_{{\rm A}{\rm C}}}{\Delta_{{\rm A}0}}+
                                  \frac{\sigma_{{\rm A}{\rm C}}}{\Delta_{{\rm C}0}}\right)\\
& & & \ds +\;\frac{\Omega_{+}}{\sqrt{3}}\;e^{-i\,\eta}
      \left(\frac{\sigma_{{\rm A}{\rm D}}\,}{\Delta_{{\rm A}0}}+
      \frac{\sigma_{{\rm B}{\rm C}}\,}{\Delta_{{\rm B}1}}-
      \frac{\sigma_{{\rm C}{\rm D}}\,}{\Delta_{{\rm C}0}}\right)\\
& & & \ds +\;\frac{\Omega_{+}}{\sqrt{6}}\;e^{i\,\eta}
      \left.\left(\frac{\sigma_{01}}{\Delta_{{\rm A}0}}-
      \frac{\sigma_{01}}{\Delta_{{\rm C}0}}\right)\right] + \rm h.c.
\end{array}
\end{equation}
where the phase $\eta$ is simply given by the difference
\begin{equation}
\eta=\eta_+-\eta_-
\end{equation}
and the detunings are defined just as~(\ref{det}) by
\begin{displaymath}
\begin{array}{ccccccc}
\Delta_{{\rm A}0} & \equiv & \omega_{{\rm A}}-\omega_0-\omega_- & ~~~ &
\Delta_{{\rm A}1} & \equiv & \omega_{{\rm A}}-\omega_1-\omega_+ \\
\Delta_{{\rm C}0} & \equiv & \omega_{{\rm C}}-\omega_0-\omega_- & ~~~ &
\Delta_{{\rm C}1} & \equiv & \omega_{{\rm C}}-\omega_1-\omega_+ \\
\Delta_{{\rm D}0} & \equiv & \omega_{{\rm D}}-\omega_0-\omega_+ & ~~~ &
\Delta_{{\rm B}1} & \equiv & \omega_{{\rm B}}-\omega_1-\omega_-
\end{array}
\end{displaymath}

When the initial atomic state is a superposition of $\ket{0}$ and $\ket{1}$,
as in our case, we can retain the terms containing $\sigma_{01}$ and
$\sigma_{10}$ only, such that the effective interaction Hamiltonian
reduces to
\begin{equation}
\tilde{\mathcal{H}}^{\rm int}=\tilde{\mathcal{H}}'_1+\tilde{\mathcal{H}}'_2
\end{equation}
where
\begin{subequations}
\label{Hprime}
\begin{eqnarray}
\tilde{\mathcal{H}}'_1 & = & \frac{\hbar\,\Omega_+\Omega_-}{4\sqrt{3}}\;e^{i\,\eta}
                             \left(\frac{\sigma_{01}}{\Delta_{{\rm A}1}}-
                                   \frac{\sigma_{01}}{\Delta_{{\rm C}1}}\right)
                             + \rm h.c.\\
\tilde{\mathcal{H}}'_2 & = & \frac{\hbar\,\Omega_+\Omega_-}{4\sqrt{3}}\;e^{i\,\eta}
                             \left(\frac{\sigma_{01}}{\Delta_{{\rm A}0}}-
                                   \frac{\sigma_{01}}{\Delta_{{\rm C}0}}\right)
                             + \rm h.c.
\end{eqnarray}
\end{subequations}
The presence of two \fpoh states with $m_F=0$ (the $\ket{\rm A}$
and $\ket{\rm C}$ states) could seem problematic. When the lasers are very
far detuned from resonance, the detunings from these two levels are
almost equal: \mbox{$\Delta_{{\rm A}0} \simeq \Delta_{{\rm C}0}$} and
\mbox{$\Delta_{{\rm A}1} \simeq \Delta_{{\rm C}1}$}. A
destructive interference between these two quantum paths suppresses
the Raman transition between the states $\ket{0}$ and $\ket{1}$
(see Eqs.~(\ref{Hprime})). In order to stimulate this
Raman transition, the detunings from levels $\ket{\rm A}$ and $\ket{\rm C}$
must be significantly different. This condition can be realized in
\rb, where the states \mbox{$\ket{\rm A} \equiv \ket{F=1,m_F=0}$} and
\mbox{$\ket{\rm C} \equiv \ket{F=2,m_F=0}$} are separated by
\mbox{$\omega_{\rm C}-\omega_{\rm A}\simeq 2\pi \cdot 800$}~MHz and the
natural line width of the level \fpoh is about \mbox{$2\pi \cdot
5.75$~MHz}. Finally, we comment the realization of the sideband excitation required by the transformation~(\ref{map}).
It requires a laser linewidth much smaller than the vibrational angular frequency,
i.e., much smaller than 10 KHz, a condition that  can be reached experimentally with
standard techniques \cite{ciro}. It is also possible to produce pulses with specific shapes that
optimize the transition probability and stimulate the transition in very short periods,
so that the gate operation time is not significantly altered.

\section{Loss and decoherence}
\label{sec:loss}

An important mechanism we have to consider in a realistic
evaluation of the performance of quantum gates, is the possibility
of loss and decoherence of the qubits during the operation.
Especially in atom chips this is an important issue, since the
cold atoms are perturbed by thermal electromagnetic fields
generated by the nearby, "hot" solid substrate.  This leads to
heating, trap loss and decoherence \cite{Henkel,Rekdal04,Varpula84}.  Since the thermal induced couplings lead to
similar timescales for all these undesired processes we consider
the loss, which was studied in largest detail theoretical and
experimentally \cite{ThLossExp}. In addition, as we have seen in the
previous section, the decoherence can be dramatically reduced by
transferring the qubits o clock states during idle periods.

Following the treatment of \cite{Henkel} one can estimate the atomic
lifetimes for a setup like shown in Fig.~\ref{fig:wires}.
Assuming gold wires we find a typical lifetime of 0.8 s due to thermally
induced spin flips and consequent loss of the qubit state. The left and
right wires, which are closest to the atoms and carry small currents, give
the largest contribution to the loss. In this configuration the best fidelity of
$\sim 98.7 \%$ is obtained for a gate operation time of $\sim 9$ ms.

Small changes in the trapping geometry can improve the fidelity:
Since a big part of the loss comes from the left and right
wires, one can fabricate them much thinner.  Reducing the thickness of the
double well wires to 50 nm would increase the lifetime to over 2
sec and increase the fidelity to $\sim$ 99.5 \% at a gate
operation time of $\sim$ 10 ms. This can be even more improved by
a little bit different wire geometry for the quadrupole wire.  A
wider but thinner quadrupole wire ($ \sim 3 \mu \mbox{m} \; \times \;50
\mbox{nm}$), and smaller left and right wires ($ \sim 0.3 \mu \mbox{m} \; \times
\;50 \mbox{nm}$) will lead to lifetimes in excess of 3 s and consequently
a better gate fidelity in the order of $\sim$ 99.7 \% at a gate
operation time of $\sim$ 11 ms.

Further improvements can be expected when optimizing the wire
geometries. Two aspects have to be taken into account.

\begin{itemize}
    \item The best achievable lifetime increases with the \emph{square}
    of the magnetic field gradient $dB/dZ$ \cite{Zhang05}.
    The present trap has a transverse gradient of 300 kG/cm. For 100 kG/cm
    and optimized wire cross section one can achieve lifetimes of up to 100 s
    \cite{Zhang05}.
    \item A second consideration is using different materials, and
    cooling the chip surface, as investigated theoretically by
    \cite{Dikov05}.  This should lead to even further improvements.
    \item Directly fabricating the wires into semiconductor chips will give
    even better thermal coupling and allow the use of high resistivity materials
    for the wires, which should improve even further the gate performance.
\end{itemize}

Overall we conclude that for optimized geometries fidelities
better than 99.9\% should be possible in realistic settings with
present day atom chip technology.

\section{Summary and conclusions}
\label{sec:Conclu}

We have presented a detailed analysis of
the implementation of a quantum phase gate with neutral
rubidium atoms on atom chips. Our analysis is quite close to the
experimental conditions and is within the reach of current
technology. We have shown how to create a double well potential near the
surface of an atom chip and studied the performance of a phase gate, using
as qubit states the vibrational states of two rubidium atoms, each sitting
in one of the two wells. The gate operation is realized through an adiabatic
modification of the potential barrier that separates the two wells.
This implementation allows selective interaction
between the vibrational states of the two atoms. We have found that, neglecting losses
due to spin flips, a fidelity of 99.9\% can be achieved in just 10.3~ms.
We have also shown that the motional qubit state can be transferred to two
hyperfine states with reduced decoherence. The estimation of loss
due to thermally induced spin flips reduces slightly the gate fidelity, and
improvements to reduce these effects have been discussed.

The results presented here are a significative improvement when compared
to an implementation using a static trap~\cite{ciro}. For a fidelity above
99\% the operation time is diminished from about 16.2~ms to
6.3~ms. The operation time could be further
reduced using non-adiabatic changes of the barrier, as recent studies
employing optimal control theory indicate~\cite{vage}, but
the excitation of the perpendicular degrees of freedom might then become
an issue.

\section{Acknowledgments}

M. A. Cirone, A. Negretti, T. Calarco and J. Schmiedmayer acknowledge financial
support from the European Union, contract number IST-2001-38863
(ACQP). T. Calarco also acknowledges financial support from the
European Union through the FP6-FET Integrated Project CT-015714 (SCALA) and a
EU Marie Curie Outgoing International Fellowship, and from the National Science
Foundation through a grant for the Institute for Theoretical Atomic, Molecular and
Optical Physics at Harvard University and Smithsonian Astrophysical Observatory.
The IDRIS-CNRS supercomputer center supported this study
by providing computational time under project number 08/051848.
This work has been done with the financial support of the LRC of
the CEA, under contract number DSM 05--33.  Laboratoire de
Photophysique Mol\'eculaire is associated to Universit\'e
Paris-Sud. We wish to thank P. Kr\"uger, J. Reichel and P.
Treutlein for useful discussions about experimental details.

\section{Appendix}
\label{sec:Appendix}

The magnetic field created by an infinitely thin and infinitely long
straight wire is
\begin{equation}
B=\frac{k_0 I}{r}
\end{equation}
where $I$ is the dc current flowing through the wire, $r$ is the distance
to the wire and $k_0=\mu_0/2\pi$ ($\mu_0$ is the magnetic permeability of the vacuum).
In the case of a wire of finite rectangular section, we average the magnetic fields
created by many infinitely thin wires lying inside the finite section. For instance,
for a dc current $I$ flowing in the positive $x$ direction through
a wire with section ($-W/2<y<W/2$, $-H/2<z<H/2$),
the magnetic field components are
\begin{equation}
B_y(y,z) = \frac{k_0I}{HW} \!\!\int_{-W/2}^{W/2}\!\!\!\!\!\!\!\!\!dy'
                               \int_{-H/2}^{H/2}\!\!\!\!\!\!\!\!\!dz'\,
                               \frac{z'-z}{(y-y')^2+(z-z')^2}
\label{by}
\end{equation}
and
\begin{equation}
B_z(y,z) = \frac{k_0I}{HW} \!\!\int_{-W/2}^{W/2}\!\!\!\!\!\!\!\!\!dy'
                               \int_{-H/2}^{H/2}\!\!\!\!\!\!\!\!\!dz'\,
                               \frac{y-y'}{(y-y')^2+(z-z')^2}
\end{equation}
In Eq.~(\ref{by}) the integration over $y'$ gives the $y$--component
\begin{equation}
B_y(y,z) \!=\! \frac{k_0I}{HW} \!\!\int_{-H/2}^{H/2}\!\!\!\!\!\!\!\!\!\!\!dz'
                               \big[\arctan\!\frac{y_-}{z-z'}-
                                    \arctan\!\frac{y_+}{z-z'}\big]
\end{equation}
where $y_{\pm} \equiv y \pm W/2$. A final integration over $z'$ gives
\begin{eqnarray}
B_y(y,z) & \!=\! & -\frac{k_0I}{HW}\Big[\,
                           z_{-}\left(\arctan\frac{z_{-}}{y_{+}}-\arctan\frac{z_{-}}{y_{-}}\right)\nonumber\\
         &       &        +z_{+}\left(\arctan\frac{z_{+}}{y_{-}}-\arctan\frac{z_{+}}{y_{+}}\right)\nonumber\\
         &       &        +\frac{y_{-}}{2}\log\frac{y_{-}^2+z_{-}^2}{y_{-}^2+z_{+}^2}
                          -\frac{y_{+}}{2}\log\frac{y_{+}^2+z_{-}^2}{y_{+}^2+z_{+}^2}\,\Big]
\end{eqnarray}
After defining $z_{\pm} \equiv z \pm H/2$, analogous calculations for the component $B_z(y,z)$ give
\begin{eqnarray}
B_z(y,z) & \!=\! & \frac{k_0I}{HW}\Big[\,
                          y_{-}\left(\arctan\frac{y_{-}}{z_{+}}-\arctan\frac{y_{-}}{z_{-}}\right)\nonumber\\
         &       &       +y_{+}\left(\arctan\frac{y_{+}}{z_{-}}-\arctan\frac{y_{+}}{z_{+}}\right)\nonumber\\
         &       &       +\frac{z_{-}}{2}\log\frac{z_{-}^2+y_{-}^2}{z_{-}^2+y_{+}^2}
                         -\frac{z_{+}}{2}\log\frac{z_{+}^2+y_{-}^2}{z_{+}^2+y_{+}^2}\,\Big]\,.
\end{eqnarray}


\begin{thebibliography}{99}
\bibitem{notedens} By insulating the wires from the substrate by an evaporated
$\mbox{SiO}_2$ layer as thin as 10 nm, relatively high current
densities like that considered in our numerical simulations can be
achieved \cite{Gro04}.
\end{thebibliography}
\end{document}